\documentclass[twocolumn,aps,showpacs,preprintnumbers,amsmath,amssymb]{revtex4}

\usepackage{graphicx}
\usepackage{dcolumn}
\usepackage{bm}

\begin{document}


\title{Charge modulation driven Fermi surface of Pb-Bi2201}

\author{L. Dudy$^1$, B. M\"uller$^1$,  B. Ziegler$^1$, A. Krapf$^1$, H. Dwelk$^1$, O. L\"ubben$^1$, R.-P Blum$^1$, V. P. Martovitsky$^2$\footnote{permanent address: Lebedev
Physical Institute, Russian Academy of Science, Leninskii pr. 53,
Moscow, 11991 Russia}, C. Janowitz$^1$, R. Manzke$^1$}

\affiliation{$^1$Humboldt-Universit\"at zu Berlin, Institut f\"ur
Physik, {Newtonstr{.15}}, D-12489 Berlin, Germany\\$^2$Department of
Low Temperature Physics, Moscow State University, Moscow 11992,
Russia}

\date{\today}

\begin{abstract}
It is well known that the (1x5) superstructure of Bi cuprate
superconductors will be suppressed due to doping with Pb.
Nevertheless, a Fermi surface map of $Bi_{2-y} Pb_y Sr_{2-x}La_x
CuO_{6+\delta}$ ($y=0.4$ and $x=0.4$) determined by angular resolved
photoemission (ARPES) revealed additional Fermi surface features.
Low energy electron diffraction and X-ray diffraction of this sample
showed no sign of any superstructure. Scanning tunneling microscopy
(STM), on the other hand, revealed two distinct modulations of the
charge density, one of (1x32) and a second of (6x6) periodicity. The
wave vectors of both modulations have been extracted and used to
simulate the corresponding Fermi surface, which is compared with the
experimental one. The origin of these modulations is discussed in
terms of dopant ordering.
\end{abstract}

\pacs{74.25.Jb, 79.60.Bm, 74.72.Hs, 68.37.Ef}

\maketitle

 Among the essential features determining the
microscopic electronic properties of high temperature (HTc)
superconductors the Fermi surface (FS) topology and the low energy
excitations at the Fermi energy ($E_F$) play a key role. Over the
last years important progress has been achieved by ARPES on various
HTc materials from which $Bi_2 Sr_2 Ca Cu_2 O_{8+\delta}$ (Bi2212)
is the most commonly studied. To investigate intrinsic features it
was found more appropriate to study crystals with partial
substitution of Bi by Pb, i.e. Pb-Bi2212, to suppress the (1x5)
superstructure in the BiO planes which could lead to unwanted
diffraction replica~\cite{ARPESrev} and mask other important
modulations of the electron system like e.g. charge
inhomogeneities/Wigner crystalization or stripes. Over the last
years it became evident that phase separations due to strong
correlations and structural and electronic inhomogeneities on an
atomic scale are essential ingredients to HTc physics. Elastic
neutron scattering experiments on $La_{1.6-x} Nd_{0.4} Sr_x Cu O_4$
can be interpreted as the occurrence of charge and spin separation
leading even to a static stripe-like arrangement near 1/8
doping~\cite{tranquada}. Also scanning tunneling microscopy (STM)
has revealed spatial variations of the quasiparticle density of
states forming a checkerboard-like structure~\cite{STMrev}.

In this publication the ramification of structural and electronic
ordering on the FS of optimally lead doped $Bi_{2-y} Pb_y Sr_{2-x}
La_x Cu O_{6+\delta}$ (Pb-Bi2201) below the scale set up by the well
known (1x5) superstructure will be discussed. The single-layer
Pb-Bi2201 compound can be regarded as a prototype material to
measure the almost undisturbed single $CuO_2$ layer, the crucial
part of HTc superconductors and therefore serves as a reference
material for the commonly studied double layered Bi2212. An
influence from the (1x5) superstructure as well as from bilayer
splitting known for Bi2212 can be ruled out~\cite{ARPESrev}. We will
show by analysis of STM patterns that two main, long range
modulations of one-dimensional and two-dimensional character exist.
Modeling the FS with inclusion of these two modulations shows
striking coincidence with the measured one by ARPES.
\begin{figure}
\includegraphics[width=86mm]{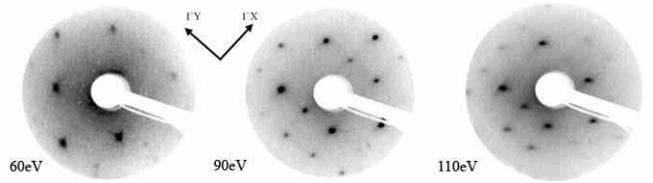}
\caption{\label{fig1} LEED-patterns of optimally doped Pb-Bi2201 for
different electron energies.}
\end{figure}

\begin{figure}
\includegraphics[width=86mm]{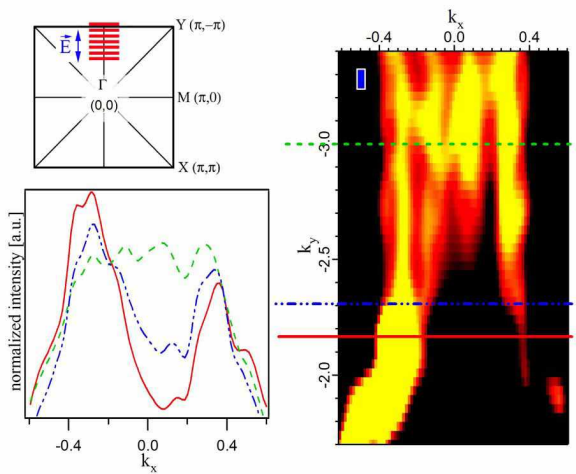}
\caption{\label{fig2} \textbf{Upper left:} First Brillouin zone of a
$CuO_2$ plane and the region of measurements. The direction of the
electrical field vector applied in the ARPES experiment is indicated
by an arrow. \textbf{Right:} Fermi suface map of $Bi_{2-y} Pb_y
Sr_{2-x} La_x Cu O_{6+\delta}$ with Pb $y=0.41$ and La $x=0.44$. The
experimental resolution is indicated by the square in the upper left
corner. The density of points was doubled in each dimension by
interpolation with a cubic spline. \textbf{Lower left:} Momentum
distribution curves taken at $E_F$ along three directions of the
Brillouin zone as indicated by the lines in the FS map. }
\end{figure}

Single crystals of Pb-Bi2201 were grown by self flux technique. The
samples were characterized by energy dispersive X-ray analysis
(EDX), AC-susceptibility measurements, LEED and X-ray diffraction
technique. According to EDX, the crystals studied had a Pb-content
of $y=0.41 \pm 0.06$ and a La-content of $x=0.44 \pm 0.10$. The
samples are optimally doped with hole concentration $n_h=0.16 \pm
0.03$ holes per Cu, $T_C^{(onset)}=30\, K \pm 2\, K$, and a
superconducting transition width  in AC-susceptibility of $\Delta
T_C^{(10\%-90\%)}=2-3\, K$. The lattice parameters measured by X-ray
diffraction are a=5.295 \AA, {b=5.322 \AA}  and c=24.448 \AA. The
LEED patterns of FIG.\ref{fig1} reveal sharp spots and no sign of a
superstructure over a sufficiently large range of electron energies.
In addition, the homogeneity of the samples was controlled by moving
the electron beam across the sample surface.

X-ray diffraction was done at a DRONE 2.0 diffractometer with
Cu$K_\lambda$-radiation and a graphite monochromator. We used high
resolution geometry with a narrow slit and a fixed pulse counter
giving an angular resolution of $0.02 ^\circ$ . The crystal was
found to be highly perfect. The only structural defect in these
crystals is a twisting along the c-axis. The twisting of adjacent
layers in the same crystal block is of about $0.1-0.3 ^\circ$,
whereas between different blocks it reaches $3-4 ^\circ$. In order
to test the crystal for even weak modulations with large periodicity
we used the strongest reflections, i.e. the (0 0 L) series. For
these reflections we measured the rocking curves along the [2 0 0]
and [0 2 0] directions. In presence of a modulation one would expect
a shift of the corresponding rocking curve maxima for different L
reflections. For example, a modulation with 32 periods would have an
angle of $0.51 ^\circ$ between (0 0 16 0) and (0 0 16 1) and an
angle of $1.01 ^\circ$ between (0 0 8 0) and (0 0 8 1). Since these
superspots have not been found it can be concluded that the crystal
is free of structural modulations.

The ARPES measurements were carried out with a Scienta SES-100
analyzer at the U125-2/10m-NIM at BESSY. All data were taken with 22
eV photon energy and a sample temperature of 25 K. The energy
resolution was 18 meV. The angular resolution was 0.2 degree. $E_F$
was determined from the Fermi-edge of an evaporated Au film. The
measured region in the first Brillouin zone is shown in
FIG.\ref{fig2}. The electrical field vector was orientated parallel
to the line $\Gamma M$ as indicated. The components of the wave
vectors are given in units in which i.e. $(k_x,k_y)=(\pi,\pi)$
denotes the X-point. We determined a FS map by integrating the
spectra in an energy window of $\pm$10 meV around $E_F$. The
resulting map is shown on the right side of FIG.\ref{fig2}. One
should note (i) the complex pattern far from what one would expect
for a single FS and (ii) the distinct anisotropy in the intensity
with respect to the symmetry direction $\Gamma M$. This is even more
obvious in the momentum distribution curves (MDC) of FIG.\ref{fig2}.
The MDC near M are more symmetric than towards $\Gamma$.

\begin{figure}
\includegraphics[width=86mm]{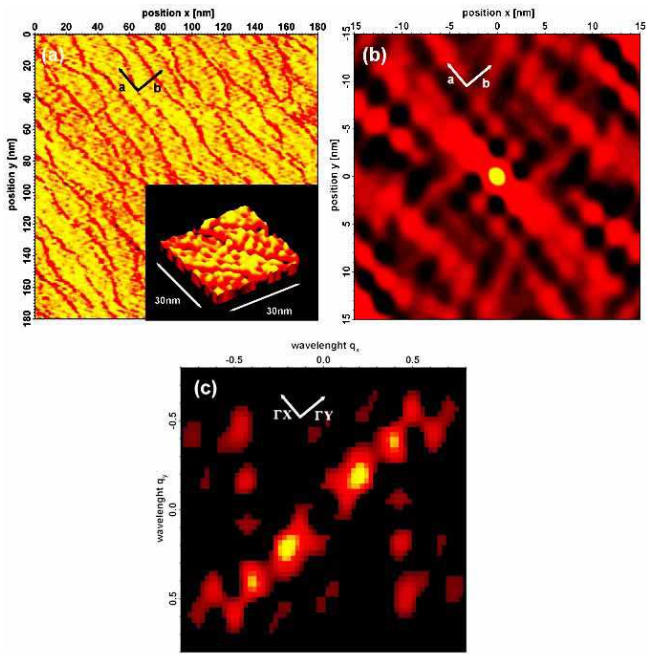}
\caption{\label{fig3} \textbf{(a)} Topological STM signal of
$Bi_{2-y} Pb_y Sr_{2-x} La_x Cu O_{6+\delta}$ with Pb $y=0.42$ and
La $x=0.40$. All measurements were done at room temperature with
$U=0.5\ V$ and $I=0.4\ nA$. In the inset an area of $30\ nm \times
30\ nm$ is magnified. \textbf{(b)} The autocorrelation $\rho_{-}
\circ {\rho(\vec u)}$
 derived for the topological STM signal shown in the inset of (a).
\textbf{(c)} The squared structure-factor $F(\vec r)^2$ as
calculated by Eq. (\ref{eq5}) from the autocorrelation function
shown in (b). The density of data points was doubled in both
dimensions by interpolation with a cubic spline. }
\end{figure}

In order to find an explanation for the complicated FS and
asymmetries observed by ARPES we analyzed in detail the corrugation
of the charge distribution at the surface. This is determined by
highly resolved STM. All measurements were performed at room
temperature with U=0.5 V and I=0.4 nA. Because of the extremely low
corrugation the noise level had to be better than 0.05 nm. A picture
obtained at these conditions is depicted in FIG.\ref{fig3} (a). Due
to the extreme two-dimensionality of the electronic structure of
Bi2201 it was found difficult to image individual atoms. At first
view only domains with an average spacing of {170 \AA} are visible
along the b-direction. The electron density inside the almost smooth
{170 \AA}-domains is also modulated, but at extremely low
corrugation. The detailed analysis described in the following will
reveal that these modulations are ordered by a (6x6) symmetry.

To obtain quantitative information on the modulations seen in STM
and their effect on the Fermi surface measured by ARPES, we first
determine the structure-factor $F(\vec q)$ from the STM pattern and
use this knowledge to simulate the FS. The result will be compared
with the experiment.

Persistent ignoring an influence of the tip, the STM signal is
proportional to the charge density $\rho(\vec r)$. It is therefore
generally possible to determine the structure-factor $F(\vec q)$
from $\rho(\vec r)$ directly. Similar to kinetic LEED theory the
structure-factor $F(\vec q)$ is given by the Fourier-transformed
charge density $\rho (\vec r)$:
\begin{equation}
{F(\vec q)}= \int d^2r \ e^{i \vec q  \vec r}\  {\rho(\vec r)}.
\end{equation}
But in order to reduce the noise level of the STM data it is more
adequate to apply the autocorrelation function of the charge density
defined as
\begin{equation}
\rho_{-} \circ {\rho(\vec u)}=\frac{1}{V} \int d^2r \ {\rho(\vec
r)}\ {\rho(\vec r + \vec u)}.
\end{equation}
With the normalization $ {\tilde\rho(\vec r)}=\rho(\vec
r)-\frac{1}{V} \int_V d\vec r \ {\rho(\vec r)}$   the
autocorrelation acts as a good natural noise-filter. The
autocorrelation $\rho_{-} \circ {\rho(\vec u)}$ of the STM data of a
30nm x 30nm region is shown in FIG. \ref{fig3}(b). The square of the
structure-factor can now be computed by use of the Wiener-Khinchin
theorem~\cite{wiener} as the absolute value of the
Fourier-transformed autocorrelation function, i.e.
\begin{equation}
\label{eq5} {F(\vec q)}^2= \left|\int d^2u \ e^{i \vec q  \vec u}\
\tilde\rho_{-} \circ {\tilde\rho(\vec u)}\right|
\end{equation}
which is shown in FIG. \ref{fig3}(c). Due to the normalization and
the finite dimension of the STM pattern we loose the information of
zero wavelength and obtain therefore no peak corresponding to the
nearly undisturbed $\rho(\vec r)$, representing the main FS and the
first order of a (1x32) modulation. On the other hand, a number of
additional peaks are detectable in $F(\vec q)^2$ which are listed in
TAB. \ref{tab1}. With highest intensity one identifies the second
order of a (1x32) modulation due to the one-dimensional ordering of
the {170 \AA}-domains. Even the third order of this modulation is
quite intense. The ordering within the flat regions of the {170
\AA}-domains yields a two-dimensional (6x6) symmetry. The weak
intensity of about 8\% is in line with the extremely low corrugation
observed by STM.

\begin{table}
\caption{\label{tab1} Position, intensity, and periodicity the peaks
found in the Fourier-transformed autocorrelation function $F(\vec
q)^2$ depicted in FIG. \ref{fig3}(c). Here the positions are given
in fractions of $\pi$, i.e. $(q_x,q_y)=(1,0)$ means the M-point.}
\begin{ruledtabular}
\begin{tabular}{ccccl}
$\pm q_x$ &$\pm q_y$ & rel. int. & periodicity & comment \\
\hline
0    &  0   &  1          & \multicolumn{2}{c} {undisturbed + 1st.order (1x32)}  \\
0,06 & -0,06&  0,42 &    16.5  b & 2nd. order (1x32)\\
0,12 &-0,12& 0,21 & 8  b & 3rd. order (1x32)\\
0,14  &  0,15& 0,08 & 6.5  a &\\
0,16 &-0,16& 0,07  &  6  b&  \raisebox{2ex}[-2ex]{(6x6)} \\

\end{tabular}
\end{ruledtabular}
\end{table}
Taking into account the periodic charge modulations found by STM the
signal of the photoelectrons near $E_F$ is given by
\begin{equation}
\label{eq1} {I(\vec k,\omega=0,T)}\simeq \int {d^2q} \ {I_0(\vec k +
\vec q, \omega = 0,T)}\  {F(\vec q)}^2.
\end{equation}
For $I_0$, the bare FS without modulations, the intensity of the
photoemission signal is the Fermi function $f(\omega,T)$ times the
spectral function $A(\vec k,\omega,T)$ (the transition matrix
element was set constant):
\begin{equation}
\label{eq6} {I_0(\vec k,\omega,T)}\simeq {f(\omega,T)} \ {A(\vec
k,\omega,T)}.
\end{equation}
We used a simplified form of the empirical spectral function $A(\vec
k,\omega,T)$ of Kordyuk et al.~\cite{kordyuk}:
\begin{equation}
{A(\vec k,\omega,T)}\simeq \frac{\Sigma''}{(\omega-{E(\vec
k)})^2+{\Sigma''}^2}.
\end{equation}
The imaginary part of the self energy is given as $
\Sigma''=\sqrt{(\alpha\omega)^2+(\beta T)^2}$. The energy and
temperature dependence is controlled by the constants $\alpha$ and
$\beta$, which were set to $\alpha=1$ and $\beta=1\,meV/K$. For the
dispersion relation $E(\vec k)$ we took a temperature independent
tight binding-like ansatz~\cite{kordyuk}
\begin{eqnarray}
\label{eq8}
{E(\vec k)}=&&{\Delta E} - 2t \left(\cos{k_x}+\cos{k_y}\right) + 4t' \left(\cos{k_x}\cos{k_y} \right)\nonumber \\
&& - 2t'' \left( \cos{2k_x}+\cos{2k_y} \right)
\end{eqnarray}
The values ${\Delta E}=0.18\ eV$ , $t=0.17\ eV$, $t'=0.015\ eV$ and
$t''=0.039\ eV$ were obtained from a fit of the dispersion curves
measured by ARPES at various Brillouin zone locations. With these
values the normalized FS-volume gives a hole concentration of
$n_h=0.5-V_{FS}=0.157 \pm 0.020$ holes per Cu which is in good
agreement with the experimental value of the studied Pb-Bi2201
crystal of $0.16 \pm 0.03$  holes per Cu.
\begin{figure}
\includegraphics[width=70mm]{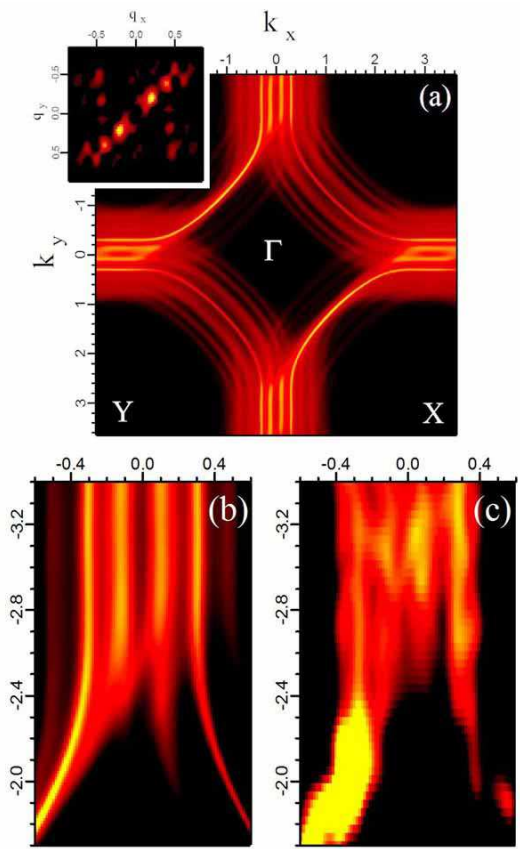}
\caption{\label{fig4} \textbf{(a)} Simulated FS of Pb-Bi2201 using
all wave vectors found in the structure-factor $F(\vec q)^2$ by STM
(see inset). \textbf{(b)(c)} Comparison of simulated FS (b) and
measured FS (c) near M. }
\end{figure}

By means of Eq. (\ref{eq6}) to (\ref{eq8}) one is now able to
compute the Fermi surface where $F(\vec q)$ includes the (1x32) and
(6x6) modulations found by STM. The intensity of the main FS
slightly broadened by the first order (1x32) was set to $I_0$ and
adjusted to the experimental intensity. This simulated FS is shown
in FIG. \ref{fig4}(a). Due to the two modulations additional FS
features appear besides the main FS causing especially around the M
$(\pi,0)$ points strong additional intensity in the simulated FS
(FIG. \ref{fig4}(b)) in accordance with the experiment (FIG.
\ref{fig4}(c)). Because of its two-dimensionality, the (6x6)
modulation leaves the FS symmetric with respect to the Brillouin
zone. The appearance of different FS intensities with respect to the
quadrants is predominantly due to the one-dimensional (1x32)
modulation. Such an asymmetrical intensity distribution has also
been observed in other experiments using Pb-Bi2201~\cite{kondo} and
therefore the (1x32) modulation gives an explanation for this
behavior.

Regarding the origin of these modulations, the following picture
emerges: While LEED and X-ray diffraction detect a perfect lattice,
STM and ARPES reveal periodic modulations. The modulations do not
reveal a distinct temperature dependence. In order to explain these
observations one should mention that LEED is not sensitive to
periodic arrangements larger than the coherence zone of about {100
\AA}. In our case the (1x32) modulation by the 1x{170 \AA} domains
is beyond that dimension and therefore not detectable by LEED. Also
modulations of low corrugation, as in our case the (6x6) structure,
are hardly detectable by LEED. Also by X-ray diffraction not all
possible periodic arrangements can be detected, namely: (i) periodic
arrangements of the Pb and Bi atoms in the otherwise perfect
lattice, and (ii) ordering of oxygen. Point (i) is due to the nearly
equivalent atomic form factors of Pb and Bi, and (ii) due to the
small atomic form factor of O. It could therefore be speculated that
a lattice solid consisting of ordered Pb, Bi and/or O could be
formed. Calculation of the thermal stability of a lattice solid in a
Pb-Bi2201 structure may be done similar as e.g. ~\cite{hoang}.  It
may be worth noting that oxygen ordering within the CuO-chains of $Y
Ba_2 Cu_3 O_x$ is quite common~\cite{widder}. It could also be
speculated that the (6x6) modulation found here exhibits a
familiarity to the nondispersive nanoscale modulations seen in
Bi2212~\cite{STMrev}. For the superconducting properties it was
recently discussed that modulations found at room temperature are
precursors to promote an ordering of holes within the $CuO_2$
plane~\cite{zhou} or modulate the superconducting pairing
interaction~\cite{bmandersen}.

In summary, we have proven that a periodic modulation of the charge
density exists and has profound influence on the Fermi surface and
generally on the electronic structure of Pb-Bi2201 observed by
ARPES.

We would like to thank the staff of BESSY and in particular G.
Reichardt for excellent support during the ARPES measurements. We
thank S. Rogaschewski for EDX and R. M\"uller for discussions. This
work has been supported by the DFG, project MA2371/3. We gratefully
thank Y.G. Ponomarev and the DAAD, program 'Ostpartnerschaften'.

\end{document}